\documentclass[twocolumn,5p,12pt]{elsarticle}
\usepackage{./aas_macros,amssymb,url}

\journal{Astroparticle Physics}

\begin{document}
\begin{frontmatter}
\title{Prospective Type Ia Supernova Surveys From Dome A}

\author[lbl]{A. Kim}
\ead{agkim@lbl.gov}

\author[cppm]{A. Bonissent}

\author[calpoly]{J. L. Christiansen}
\author[cppm]{A. Ealet}
\author[cal]{L. Faccioli}
\author[penn]{L. Gladney}

\author[lbl]{G.~Kushner}

\author[cal]{E. Linder}

\author[fnal]{C. Stoughton}

\author[tam,ccaa]{L. Wang}

\address[lbl]{Lawrence Berkeley National Laboratory,
    Berkeley, CA 94720}
\address[cppm]{Centre de Physique des Particules de Marseille, Marseille, France}
\address[calpoly]{California Polytechnic State University, San Luis Obispo, CA, 93407}

\address[cal]{University of California, Berkeley, CA 94720}
\address[penn]{University of Pennsylvania, Philadelphia, PA 19104}

\address[fnal]{Fermi National Accelerator Laboratory, Batavia, IL 60510}
\address[tam]{Texas A\&M University, College Station, TX  77843}
\address[ccaa]{Chinese Center for Antarctic Astronomy, Nanjing, China}

\begin{abstract}
Dome A, the highest plateau in Antarctica, is being developed as a site for an astronomical
observatory.  The planned telescopes and instrumentation and the unique site
characteristics are conducive toward Type Ia supernova surveys for cosmology.  
A self-contained search and survey over five years can yield a spectro-photometric 
time series of $\sim$1000 $z<0.08$ supernovae.  These can serve to anchor the 
Hubble diagram and quantify the relationship between luminosities and heterogeneities 
within the Type Ia supernova class, reducing systematics.  Larger aperture 
($\gtrsim$4-m) telescopes are capable of discovering supernovae shortly after 
explosion out to $z \sim 3$.  These can be fed to space telescopes, and can isolate 
systematics and extend the redshift range over which we measure the expansion 
history of the universe. 
\end{abstract}

\end{frontmatter}

\section{Introduction}
Dome A, the highest plateau in Antarctica, is being considered as a site for
an optical-to-infrared observatory
\cite{2009astro2010S.308W,2009PASP..121..976S}.  There are several attributes that make Dome A attractive for 
Type Ia supernova (SN Ia) observations compared to temperate sites. 
The boundary layer is at $\sim$20 m, above which there is a median optical seeing of 0.3" 
(infrared seeing of 0.2'').  During 
the long winter night, there are no disruptions over 24 hours due to 
weather.  The atmospheric column density is relatively low. The sky surface brightness in $K_{\rm dark}
$, from 2.27--2.45 $\mu$m is expected to be at $\sim$100 $\mu\mbox{Jy arcsec}^{-2}$;  OH emission drops out 
in these wavelengths
while sky and telescope thermal emission is suppressed compared to temperate sites
as is seen at the South Pole \citep{1999ApJ...527.1009P} and Dome C \citep{2005PASA...22..199B}.

Dome A has its disadvantages as well.  The long day interrupts the longer-term 
observations necessary for tracking supernova flux evolution.  The polar latitude limits the 
area of sky accessible by a telescope. Auroral activity produces high UV to $B$ band 
sky emission with non-uniform spatial structure.
There are technical hurdles as well, including the remoteness of the site, power, and 
the transfer of data out of the site, avoiding mirror condensation, and achieving 
the excellent seeing with the observatory structures. 

China is moving forward with developing the site \citep{2009PASP..121..174Y}.  In the (austral) 
summer of 2008-2009 a summer station was constructed.  
In the next few years, the three 0.5-m Antarctic Schmidt Telescopes (AST3) will be installed.  A 1-m class
pathfinder is planned to characterize the site and develop the necessary technical knowhow; the intent
is to eventually build a 4-10 m-class telescope.

In this paper, we explore possible supernova surveys that can be performed by these telescopes.
In \S\ref{lowz:sec} the search and follow-up of low-redshift supernovae are discussed.
In \S\ref{highz:sec} the possibilities for a high-redshift search are explored.  We summarize
our findings in \S\ref{conclusions:sec}.

\section{Low-Redshift Supernovae}
\label{lowz:sec}
A nearby sample is essential when using Type Ia supernovae as distance indicators to measure 
the expansion history of the universe \cite{2006PhRvD..74j3518L}. The low-redshift anchor 
has strong leverage in fits of cosmological parameters. 
In addition, the accessibility of  accurate and precise photometric and
spectroscopic measurements of these high flux objects makes them an 
important set with which to gain a quantitative understanding of the supernova luminosity function 
and its sub-populations.

There are currently relatively few discovered and followed supernovae in the nearby
Hubble flow at $z\approx 0.03$--0.1 as compared to higher redshifts, and there are several programs designed to bolster their numbers.  The Nearby Supernova Factory \cite{2002SPIE.4836...61A}, for example,
targets the range $0.03<z<0.08$ as a sweet spot \cite{2006PhRvD..74j3518L} where peculiar 
velocities are relatively small compared to the cosmological redshift and the deviation 
from the linear Hubble law is sensitive to the current deceleration parameter $q_0$ but 
not to individual cosmological components.  
At the current level of systematics, the Hubble 
diagram would be anchored satisfactorily by the 180 well-observed nearby supernovae we take 
to exist by 2011.  Thus the main strength of the Dome A low-redshift supernovae is its ability to study 
the supernova properties themselves fulfilling the
important role of mapping out systematic effects with a large sample. 
This includes 
finding subpopulations that may be relatively rare today though more common at higher redshifts. 
If we seek to distinguish subclasses at the 0.02 mag level in absolute magnitude, in the presence 
of an intrinsic scatter of 0.15 mag, this statistically requires 50 supernovae.  To detect a subclass down to 
10\% of the total population we would need a full set of 500 supernovae in that redshift range. 

These two requirements of low enough redshift to allow thorough observation and high enough 
numbers to allow in-depth characterization drive the survey design. For the purposes of this note 
we consider a survey over $0.03<z<0.08$.  A challenge to be faced for the low-redshift supernova 
discoveries is scanning the large sky area necessary to generate a significant sample. 
The supernova rate at $z\sim0.1$ is $2.93 \times10^{-5}$ $ \mbox{SNe Mpc}^{-3} h^3_{70} 
\mbox{yr}^{-1}$ \cite{2008ApJ...682..262D} giving an observed rate $d^2N/(d\Omega\,dt)= 2.0\,\delta z$
$\mbox{deg}^{-2}\mbox{yr}^{-1}$ at $z\sim 0.07$.  The available area of sky with low airmass 
and Galactic dust absorption is 
limited at Dome A.  At latitude 80.37$^\circ$ S, the airmass of a field with declination $\delta$ is
$\csc{(\delta)}$ to first order, and 
a significant fraction of sky at the south equatorial pole is covered by the Milky Way.

The useful sky available during the Antarctic winter at differing Galactic extinction can be
determined using the 
dust maps of Schlegel, Finkbeiner, and Davis (1998) \cite{1998ApJ...500..525S}.  
For 
$E(B-V)<0.05$ over the winter, 
the available sky area ranges from 1600--2900 deg$^2$ with airmass  $\chi<1.5$ that
expands to 2900--4400 deg$^2$ with $\chi<2$.  Extending to $0.05<E(B-V)<0.2$ further
increases the total sky area to 7300--7800 deg$^2$.
For the purposes of this paper we adopt exposure times for two 4000 deg$^2$ areas, one with
with $E(B-V)=0.05$ and the other with $E(B-V)=0.2$ . In three months 100--200 new supernova explosions are expected
from $0.03<z<0.08$.  

AST3 is comprised of three 0.5-m telescopes each equipped with an optical imager with a 4.5 deg$^2$ 
field of view with 1.0" pixels with readout much shorter than integration times.  The 4000-8000 deg$^2$ is covered with 890--1780 pointings.  The supernovae are sparse with less than 
one active (within two months of explosion) target per imager footprint on average.  

In the following calculations we use the updated supernova template spectral time series of Hsiao et al.\ (2007) 
\cite{2007ApJ...663.1187H} with the
mean SN Ia absolute $B$ magnitude of $-19.46+5\log{(h/0.6)}$ found by Richardson et al.\ (2002) 
\cite{2002AJ....123..745R}, where $h$ is the Hubble constant in units of 100 km/s/Mpc.

We assume an 80\% system throughput for imaging and 60\% for spectroscopy.  The optical to NIR sky brightness is taken to be that of 
CFHT at full moon, as is atmospheric absorption, since Dome A has similar sky to temperate
sites at these wavelengths.  The sky brightness from 2.27--2.45 $\mu$m 
is taken to be 100 $\mu$Jy/arcsec$^2$, the value measured at Dome C \citep{2005PASA...22..199B}.
A seeing of $0.3" (\lambda/0.5\mu\mbox{m})^{-0.2}$ added quadratically with the 
diffraction limit and pixelization yields the effective PSF used for 1 FWHM aperture photometry.  For the case of AST3 the 
effective point spread function is dominated by the undersampled pixels.
Supernova signal-to-noises and exposure times are calculated with SNAPsim, a
simulation package of astronomical observations developed by the Supernova Acceleration Probe collaboration
\cite{2002SPIE.4835..146A}. 

\begin{figure}
\begin{center}
\includegraphics[scale=0.48]{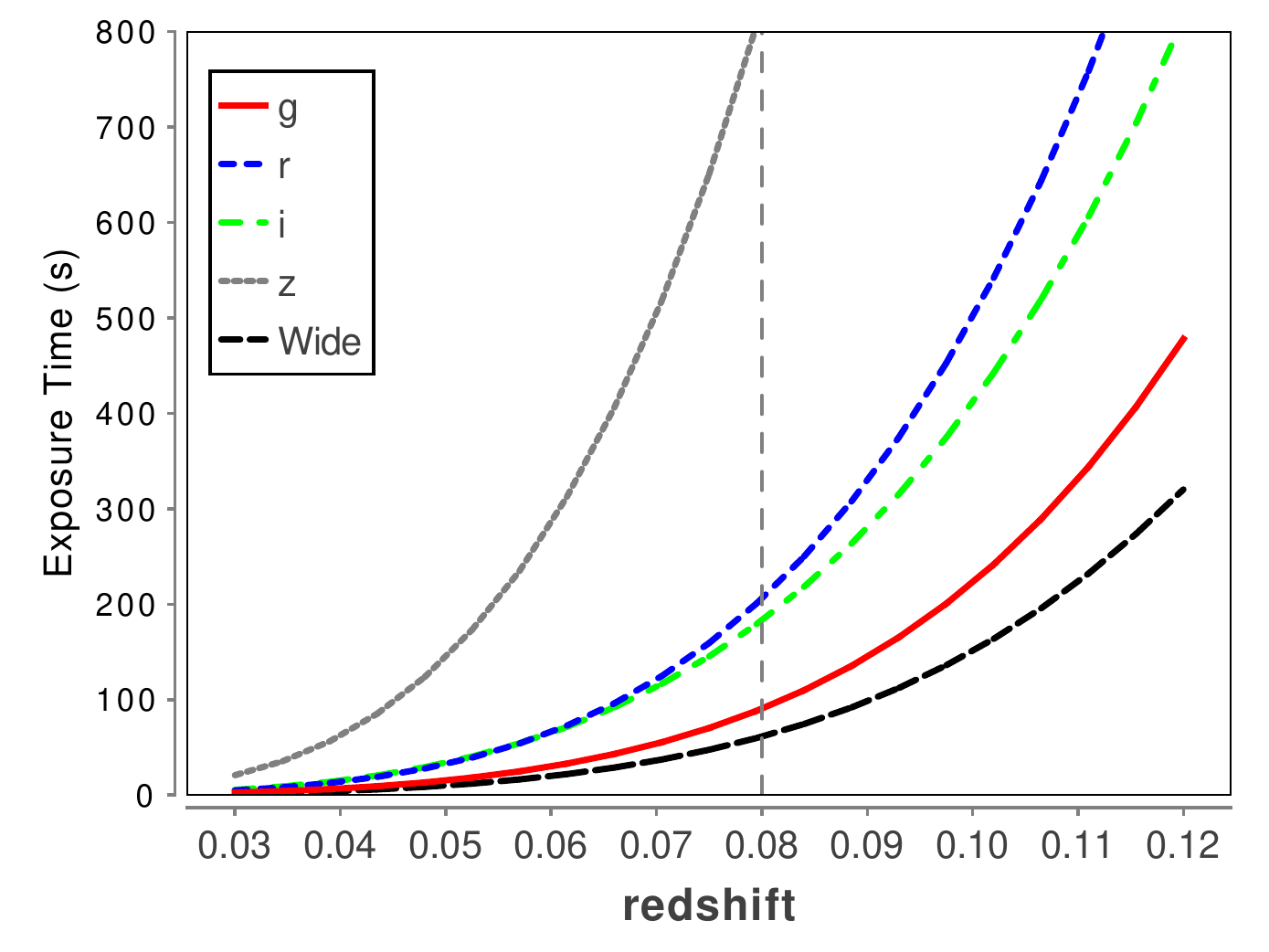}
\includegraphics[scale=0.48]{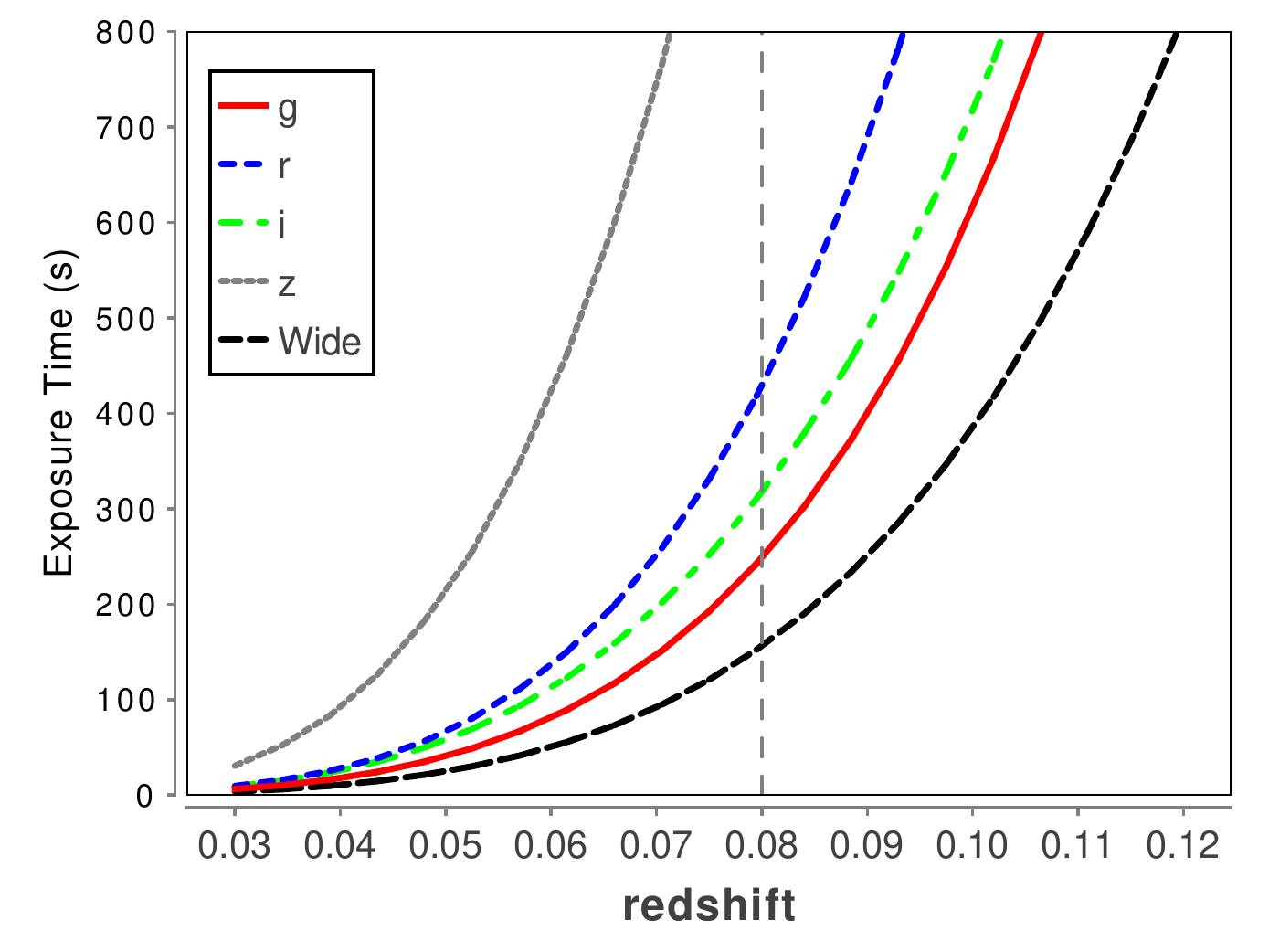}
\end{center}
\caption{The exposure times required to obtain $S/N=5$ five days after explosion for a canonical
supernova at AST3 using the Sloan $g$, $r$, $i$, $z$ filters and a wide filter extending from 0.4-0.65
$\mu$m.  The figure on the left corresponds to a Galactic extinction of $E(B-V)=0.05$
and the right to $E(B-V)=0.2$. \label{ast3:fig}}
\end{figure}

\begin{figure}
\begin{center}
\includegraphics[scale=0.48]{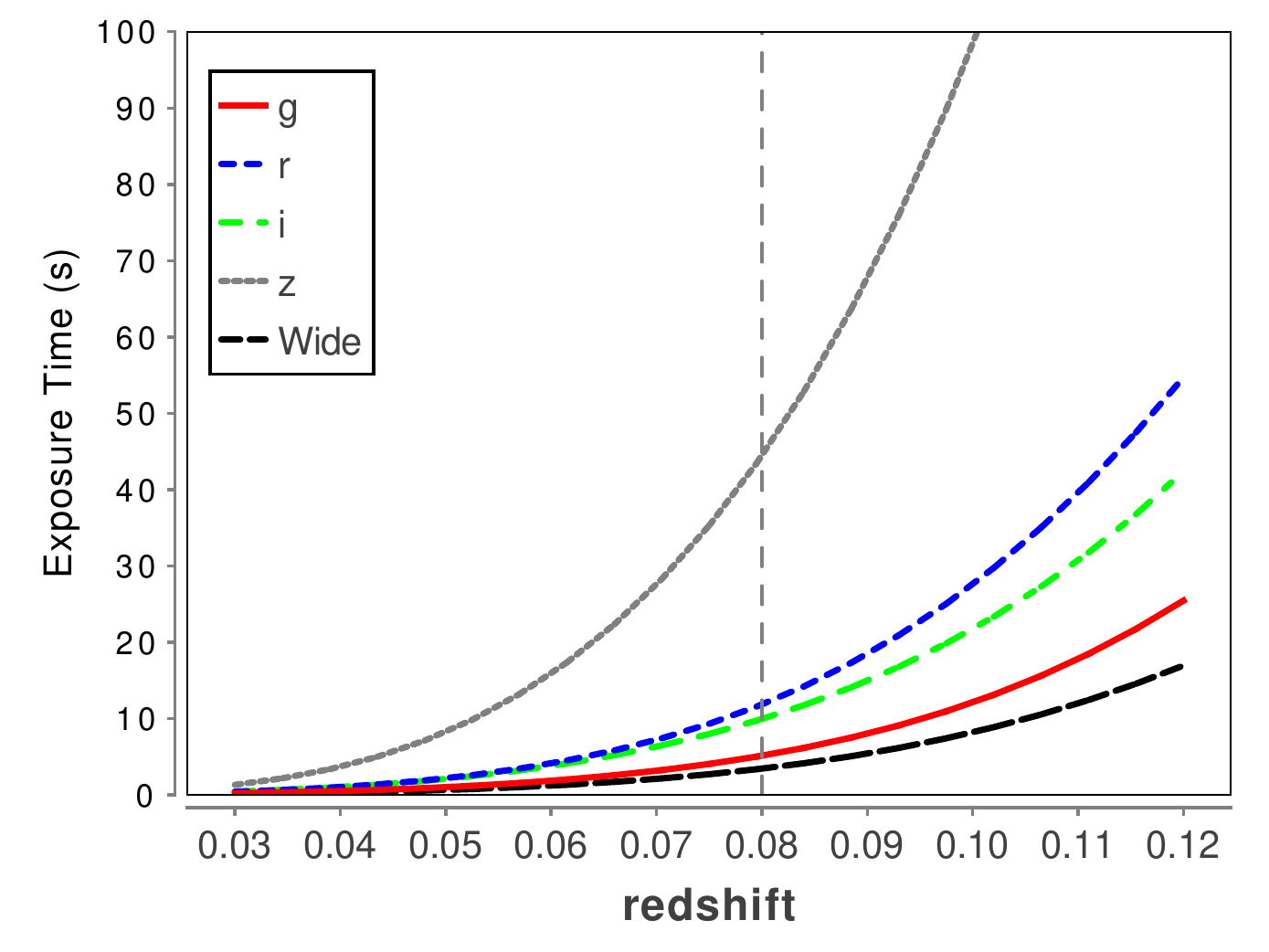}
\includegraphics[scale=0.48]{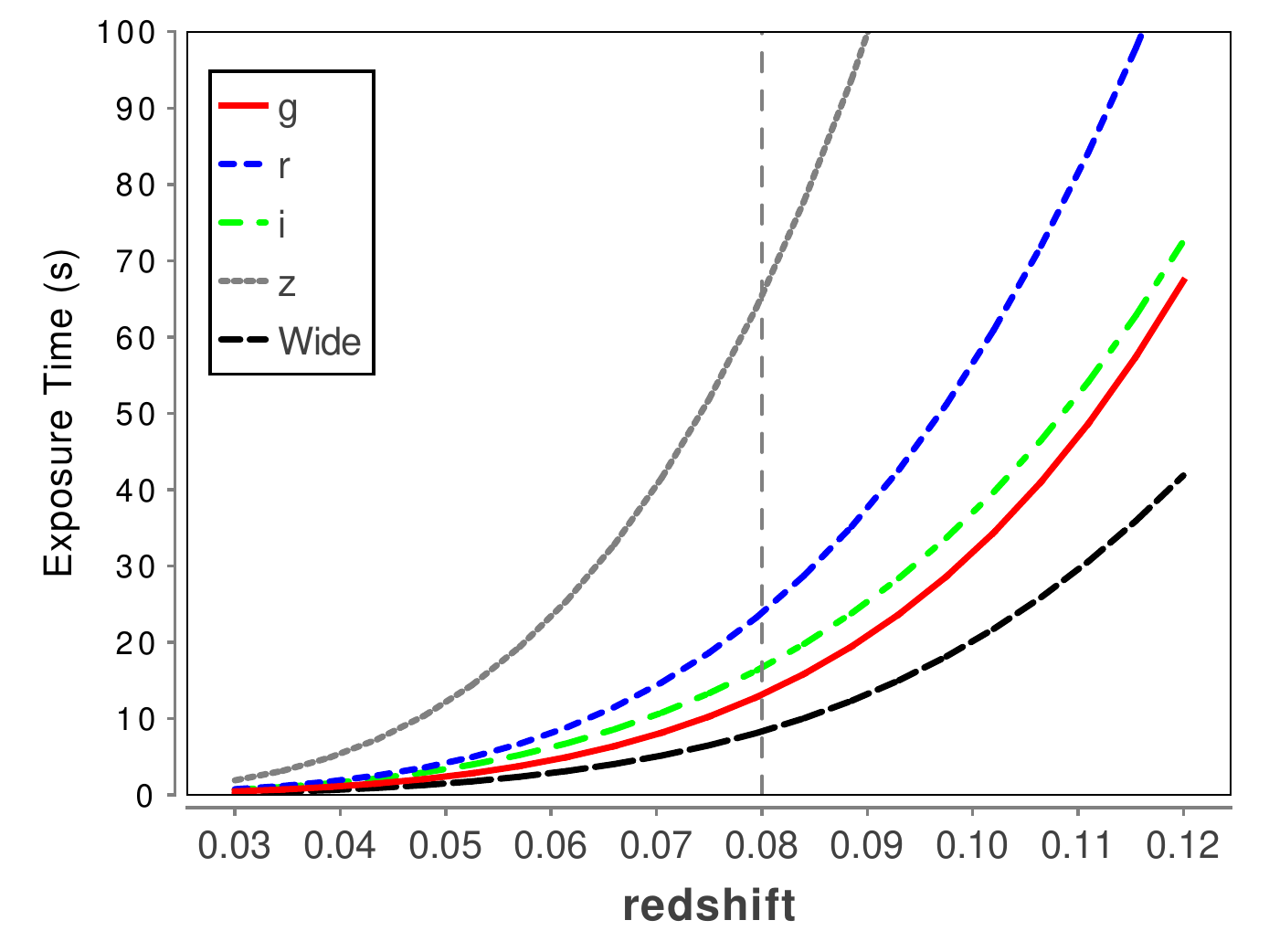}
\end{center}
\caption{The exposure times required to obtain $S/N=5$ ten days after explosion for a canonical
supernova at AST3 using the Sloan $g$, $r$, $i$, $z$ filters and a wide filter extending from 0.4-0.65
$\mu$m.  The figure on the left corresponds to a Galactic extinction of $E(B-V)=0.05$
and the right to $E(B-V)=0.2$.   Note that these plots have different ordinate ranges compared to those in Figure~\ref{ast3:fig}.\label{ast3_late:fig}}
\end{figure}

The requirement for the early discovery of supernovae is given as $S/N=5$ five rest-frame days after
explosion in the Hsiao et al.\ (2007) \cite{2007ApJ...663.1187H} model.
Figure~\ref{ast3:fig} shows the required 
exposure times for fields with $E(B-V)=0.05$ and $E(B-V) =0.2$.  For a goal of $z_{\rm max}=0.08$,
the required times with the wide (0.4-0.65 $\mu m$) filter are 60 and 150 seconds while the narrower Sloan bands require longer exposures.   A 0.5-m class telescope, even if equipped with severely undersampled pixels, can provide sufficient signal-to-noise for early discovery of supernovae at low redshift.
Covering the 8000 deg$^2$ takes $\sim$52 hours of exposure plus overhead.  If 16 of 24
hours are available for integration time, then it would take the three telescopes together a little over a day to observe the 8000 deg$^2$, enabling a 1-day search cadence.  Figure~\ref{ast3_late:fig} shows
how the
required exposure times are greatly reduced if the discovery phase is relaxed to ten restframe days after explosion.

\begin{figure}
\begin{center}
\includegraphics[scale=0.48]{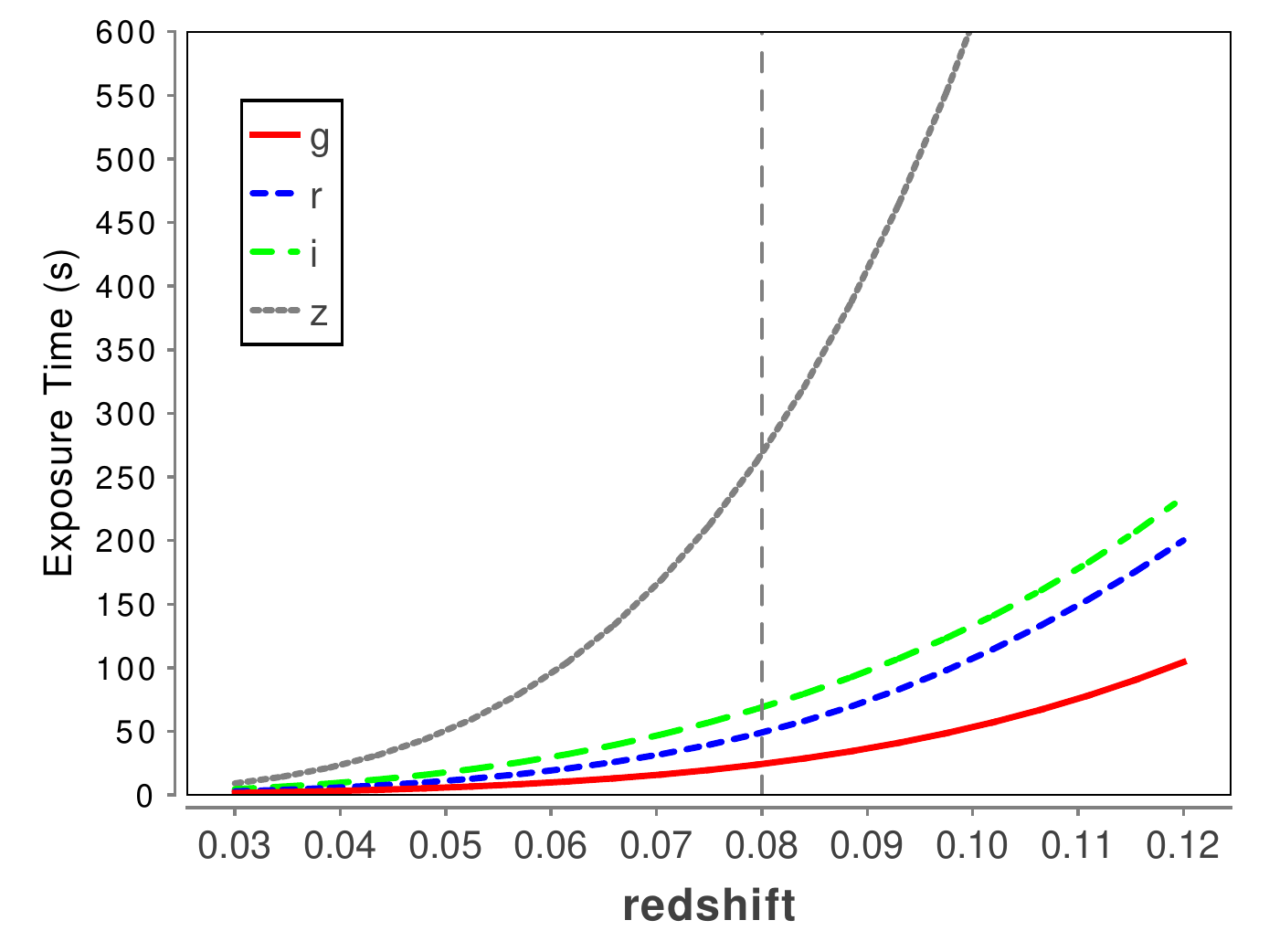}
\includegraphics[scale=0.48]{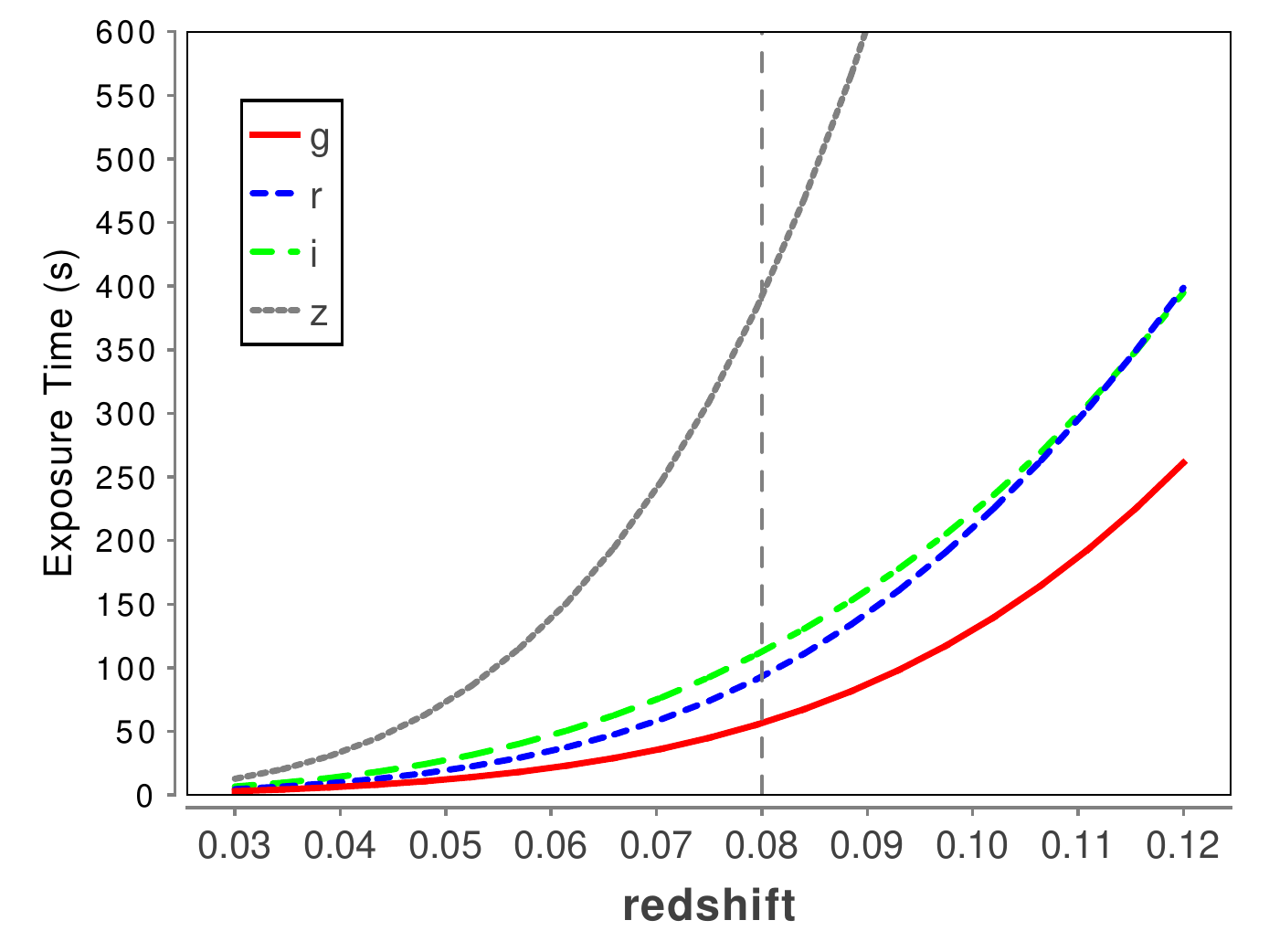}
\end{center}
\caption{The exposure time required to obtain $S/N=25$ at maximum light for a canonical
supernova at AST3 in each of the Sloan $g$, $r$, $i$, and $z$ filters.  The figure on the left corresponds to a 
Galactic extinction of $E(B-V)=0.05$
and the right to $E(B-V)=0.2$. \label{ast3_peak:fig}}
\end{figure}

As an overall requirement for the light curve we adopt a $S/N=25$ at peak brightness in each band, used to calculate accurate colors.
The needed exposure times for the Sloan bands are shown in Figure~\ref{ast3_peak:fig}.
The per-band exposure times are between those necessary for five- and ten-day post-explosion discovery.
Meeting the requirements in $g$, $r$, and $i$ requires 145~s and 260~s integrations for $E(B-V)=0.05$
and 0.2 respectively.   With the three telescopes used in a rolling search, the light curves could be sampled every 36 hours.  Although most of the images will not have an active supernova such a survey requires no active exposure time nor telescope pointing decisions so it runs passively with off-line data analysis.  
(Moreover it would pick up other transients and build deep images of the sky.) 

\begin{figure}
\begin{center}
\includegraphics[scale=0.48]{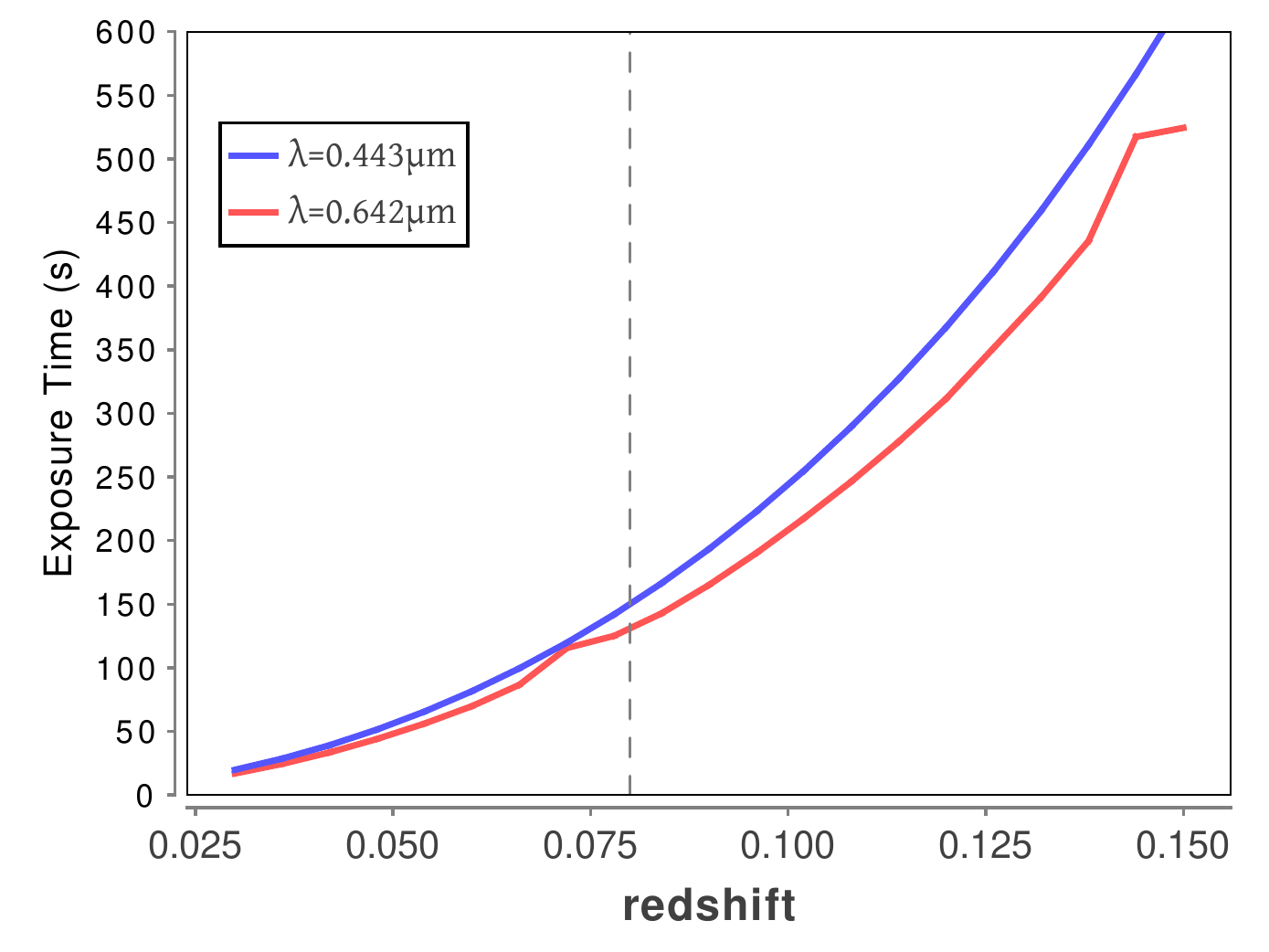}
\includegraphics[scale=0.48]{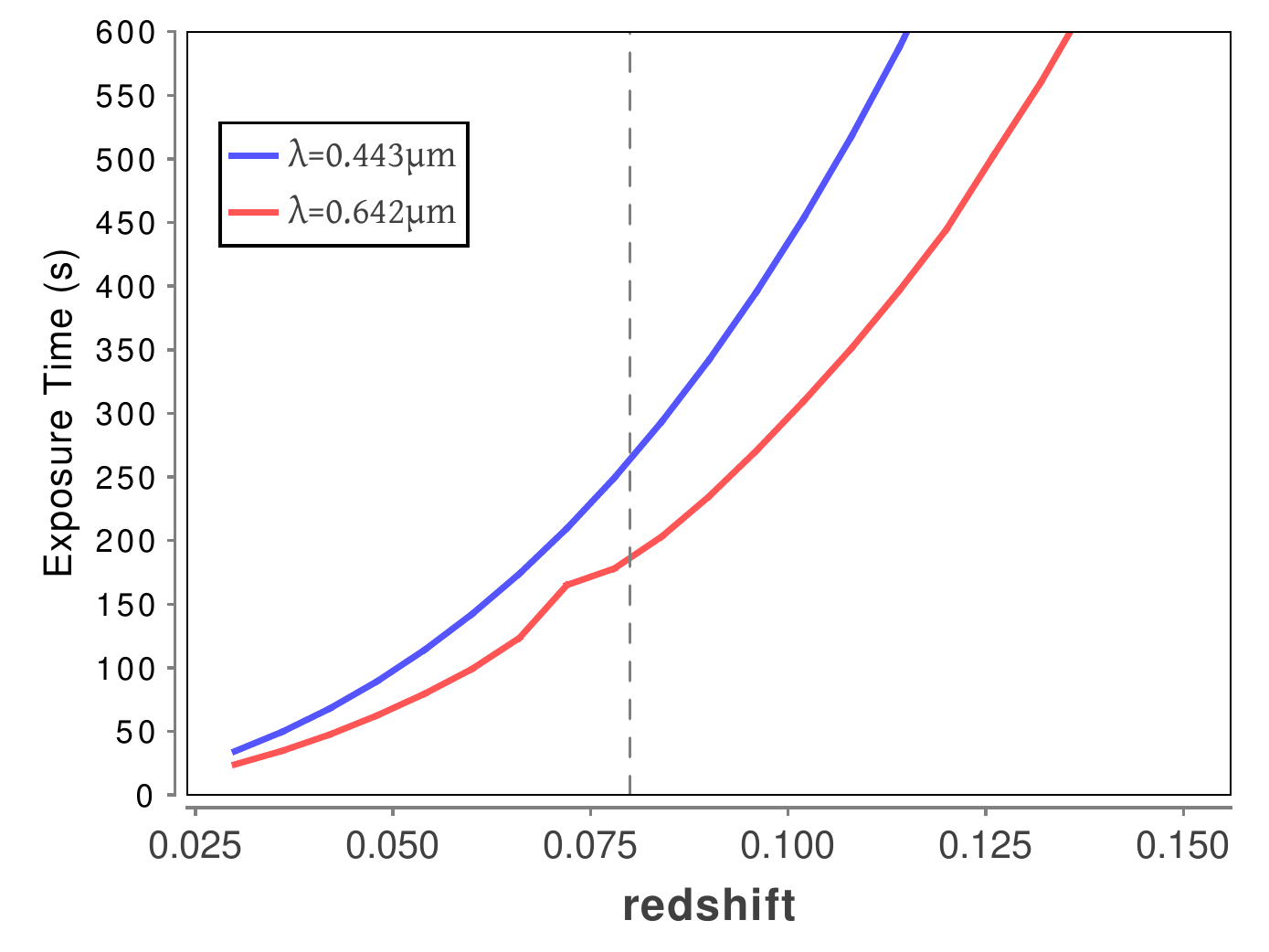}
\end{center}
\caption{The exposure time required to obtain $S/N=25$ at maximum light in a $\lambda/\delta 
\lambda=150$ bin  at rest-frame $0.443$ and $0.642$ $\mu$m for a canonical
supernova using a 1-m telescope at Dome A with an integral-field-unit spectrograph.
The figure on the left corresponds to a Galactic extinction of $E(B-V)=0.05$
and the right to $E(B-V)=0.2$. \label{onem_spec:fig}}
\end{figure}

The 1-m pathfinder telescope can be used for spectroscopic followup.  We 
consider an integral-field-unit spectrograph (IFU) that critically samples the 0.3" seeing.  Its field
of view is large enough to cover the supernova and several field sources needed for calibration.
The spectrograph covers the full optical spectrum from $0.35<\lambda<1$ $\mu$m with resolution $R \gg 75$ while not being detector-noise dominated.   
(An $R\sim75$ is a minimum for supernovae while a higher resolution is important for site testing.)  We examine a peak-brightness exposure time requirement of $S/N=25$ in
a $\lambda/\delta \lambda=150$ bin at rest-frame $0.443$ and $0.642$ $\mu$m; this
is the level of accuracy to detect heterogeneity in the spectral line ratios found by 
Bailey et al.\ (2009) \cite{2009A&A...500L..17B} that calibrate supernova distances beyond
standard accuracies.  Exposure times as a function of
redshift are shown in Figure~\ref{onem_spec:fig}.  Supernovae at $z=0.08$ in low-extinction fields require $\sim$150~s
 exposure times and 
the high-extinction fields need $\sim$250~s.  Observing 200 supernovae with these 
exposure times requires $\sim 11.1$ hours of integration time, short enough to allow daily spectro-photometric 
sampling with the fixed exposure time.  
In a more realistic implementation, exposure times would be tuned to the actual seeing, targeted signal-to-noise, and magnitude at the date of observation, which may be brighter or fainter than our adopted number depending on supernova redshift and phase.

\begin{figure}
\begin{center}
\includegraphics[scale=0.5]{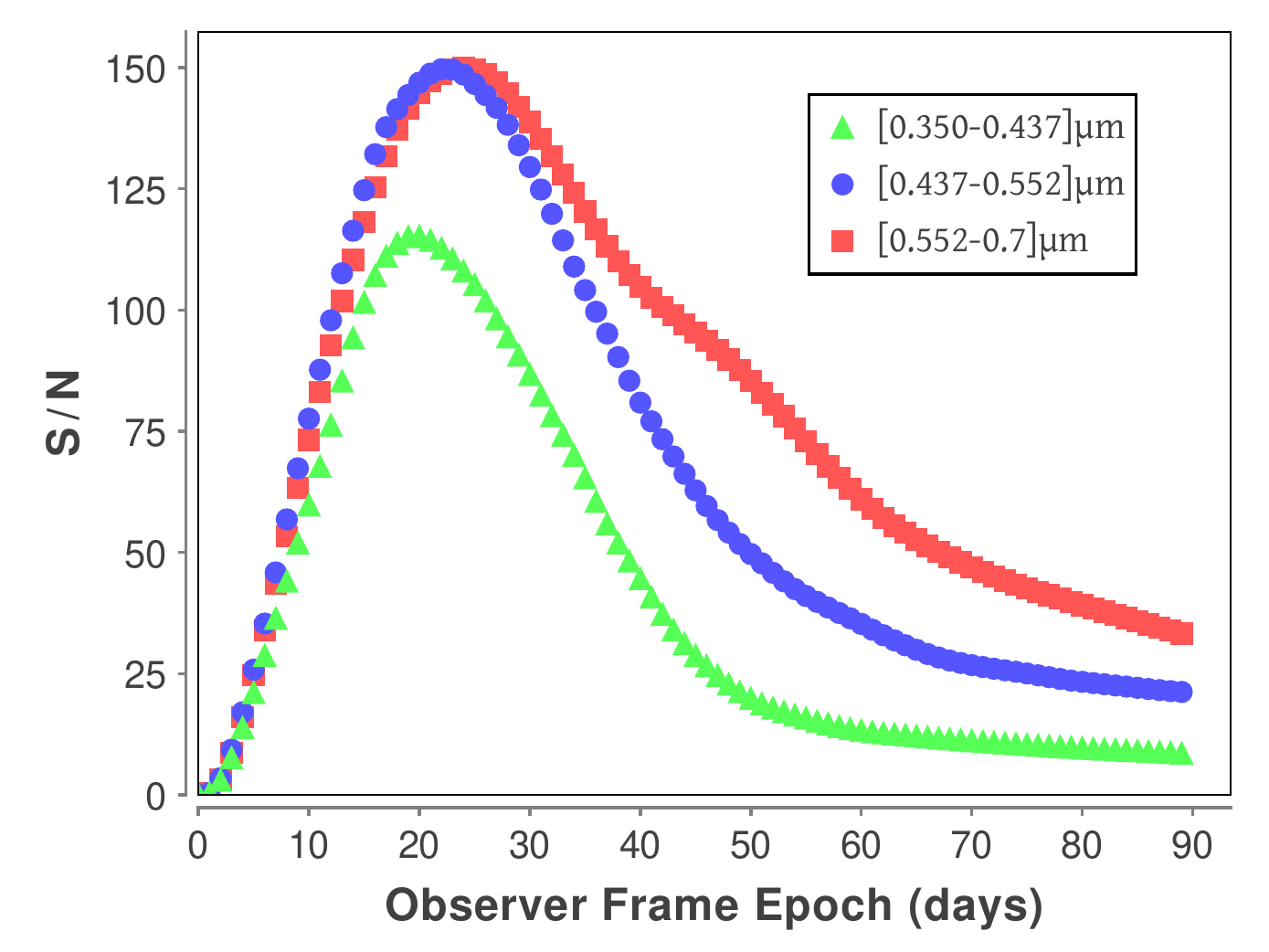}
\end{center}
\caption{The signal-to-noise in three synthetic bands for a supernova at $z=0.08$: from
[0.35--0.437], [0.437--0.552], and [0.552--0.7] $\mu$m based on
the conditions given in Figure~\ref{onem_spec:fig}. The light curves are based on spectra
taken every 24 hours with an exposure of 150~s.\label{onem_ifu_effphot:fig}}
\end{figure}

Daily spectro-photometric spectra can be integrated to produce light curves in synthetic
photometric bands.  Figure~\ref{onem_ifu_effphot:fig} shows the effective light curve
signal-to-noises
in [0.35--0.437], [0.437--0.552], and [0.552--0.7] $\mu$m bands for 150~s exposures of a $z=0.08$ supernova.
The exposure time, defined by the requirement to measure spectral ratios at peak brightness,
generates light curves with excellent signal-to-noise throughout months of the supernova's evolution.

\begin{figure}
\begin{center}
\includegraphics[scale=0.5]{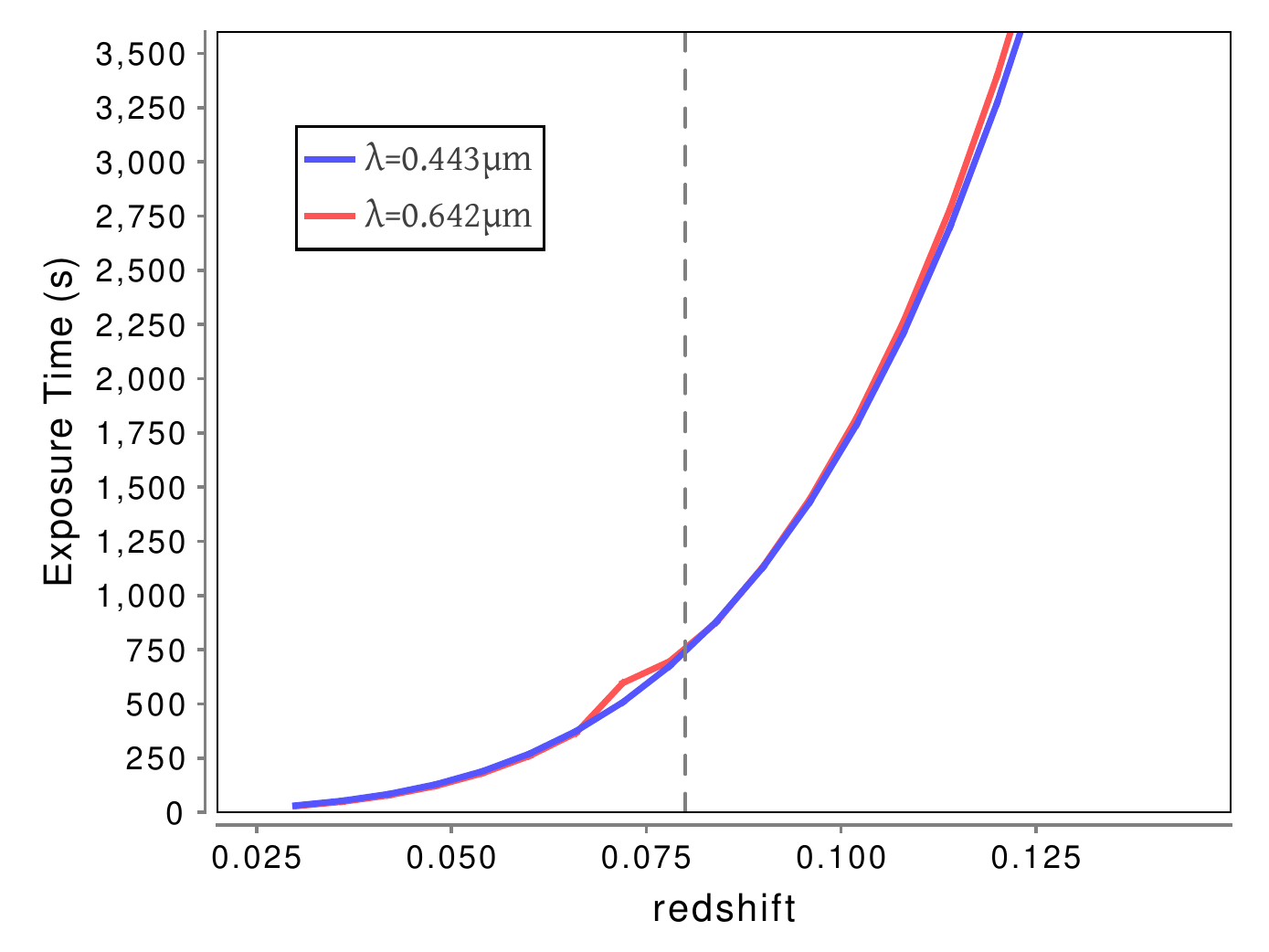}
\end{center}
\caption{The exposure time required to obtain $S/N=25$ at maximum light in a $\lambda/\delta 
\lambda=150$ bin at rest-frame 0.443 and 0.642 $\mu$m for a canonical
supernova using a 1-m telescope at Dome A with slitless spectrososcopy split into
0.35--0.5 $\mu$m and 0.5--0.7 $\mu$m channels
for $E(B-V)=0.05$. \label{onem_grism:fig}}
\end{figure}

The low targeted redshifts, the excellent seeing at Dome A, and the low wavelength resolution needed to 
measure broad spectroscopic features result in noise that is strongly source-limited for the
IFU spectrograph.  From a 
statistical standpoint this makes slitless spectroscopy a possible alternative.  As an example, consider
two grism channels, one from 0.35--0.5 $\mu$m and the second from 0.5--0.7 $\mu$m
with $R=75$.  The wavelength ranges are tuned to give matching exposure times for the two targeted wavelengths.  
The exposure times necessary for the same requirements used for the IFU are shown in Figure~\ref{onem_grism:fig}; for $z=0.08$ the time is
750~s.   In the lower redshift range 
the exposure times are the same as with the IFU, while at higher redshift the sky 
contribution dominates and exposure times increase. Slitless spectroscopy also requires
longer exposures at supernova phases off peak brightness, where again the sky is the dominant
background.

\begin{figure}
\begin{center}
\includegraphics[scale=0.5]{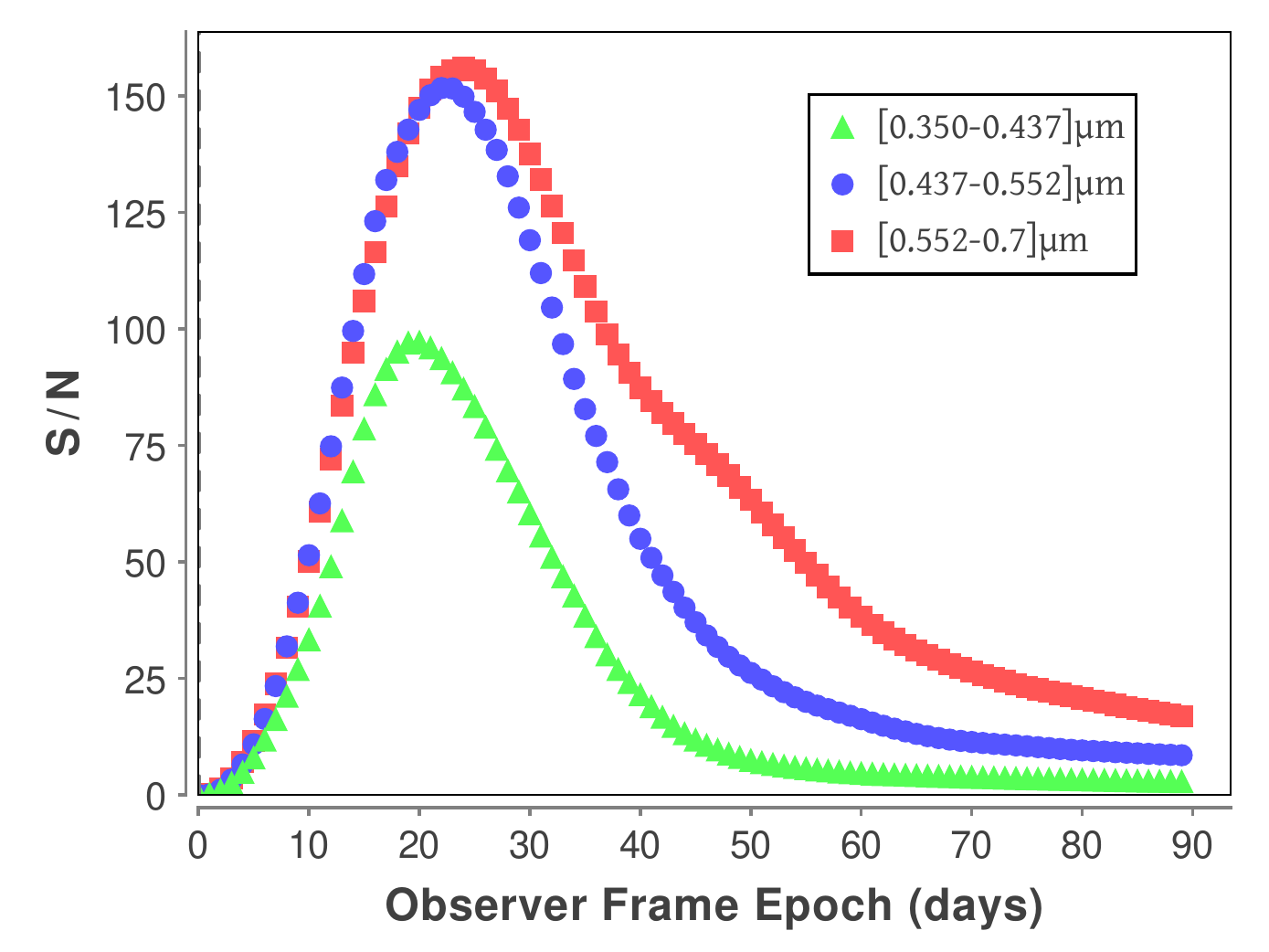}
\end{center}
\caption{The signal-to-noise in three synthetic bands for a supernova at $z=0.08$: from
[0.35--0.437], [0.437--0.552], and [0.552--0.7] $\mu$m based on
the conditions given in Figure~\ref{onem_grism:fig}. The light curves are based on spectra
taken every 24 hours with an exposure of 750~s.\label{onem_grism_effphot:fig}}
\end{figure}

A wide-field slitless spectroscopic camera with a cadenced survey can generate
spectroscopic time-series of all the supernovae and other bright transients in the field 
without the need for triggering.  This deterministic survey requires no human intervention 
during data collection, and thus is attractive for a site such as Dome A.
Synthetic photometry can be produced from the spectrum.  Figure~\ref{onem_grism_effphot:fig}
shows the signal-to-noise over daily-sampled synthetic light curves obtained for a $z=0.08$, $E(B-V)=0.05$ supernova with a 750~s integration.   Given the respective exposure times set by the requirement at peak brightness, slitless spectroscopy is qualitatively similar to the IFU (Figure~\ref{onem_ifu_effphot:fig}) except at early and late phases when the supernova is faint.  Slitless
spectroscopy is less effective in detecting  possible signatures of SN Ia heterogeneity at early 
and plateau phases \cite{2007ApJ...671.1084S,2003LNP...635..203H}.
Such a focal plane faces a huge cost and technological hurdles considering the number of pixels
needed for critical sampling; an imager with 0.15" pixels requires 0.578 Gpix to cover one square degree. 

Another issue of interest concerns supernova observations in the near-infrared. 
Supernova $J$ and $H$ peak magnitudes without extinction or
light-curve shape correction are found to have similar RMS ($\sim
0.15$) as those of corrected optical magnitudes
\citep{2004ApJ...602L..81K,2008ApJ...689..377W}.
This smaller intrinsic dispersion in the infrared is
predicted by Kasen (2006) \cite{2006ApJ...649..939K}.  Dust absorption uncertainties
are significantly reduced as well.  Expanding the set of supernovae with NIR wavelengths
can help establish whether the near-infrared provides a viable alternative way to measure distances.

\begin{figure}
\begin{center}
\includegraphics[scale=0.5]{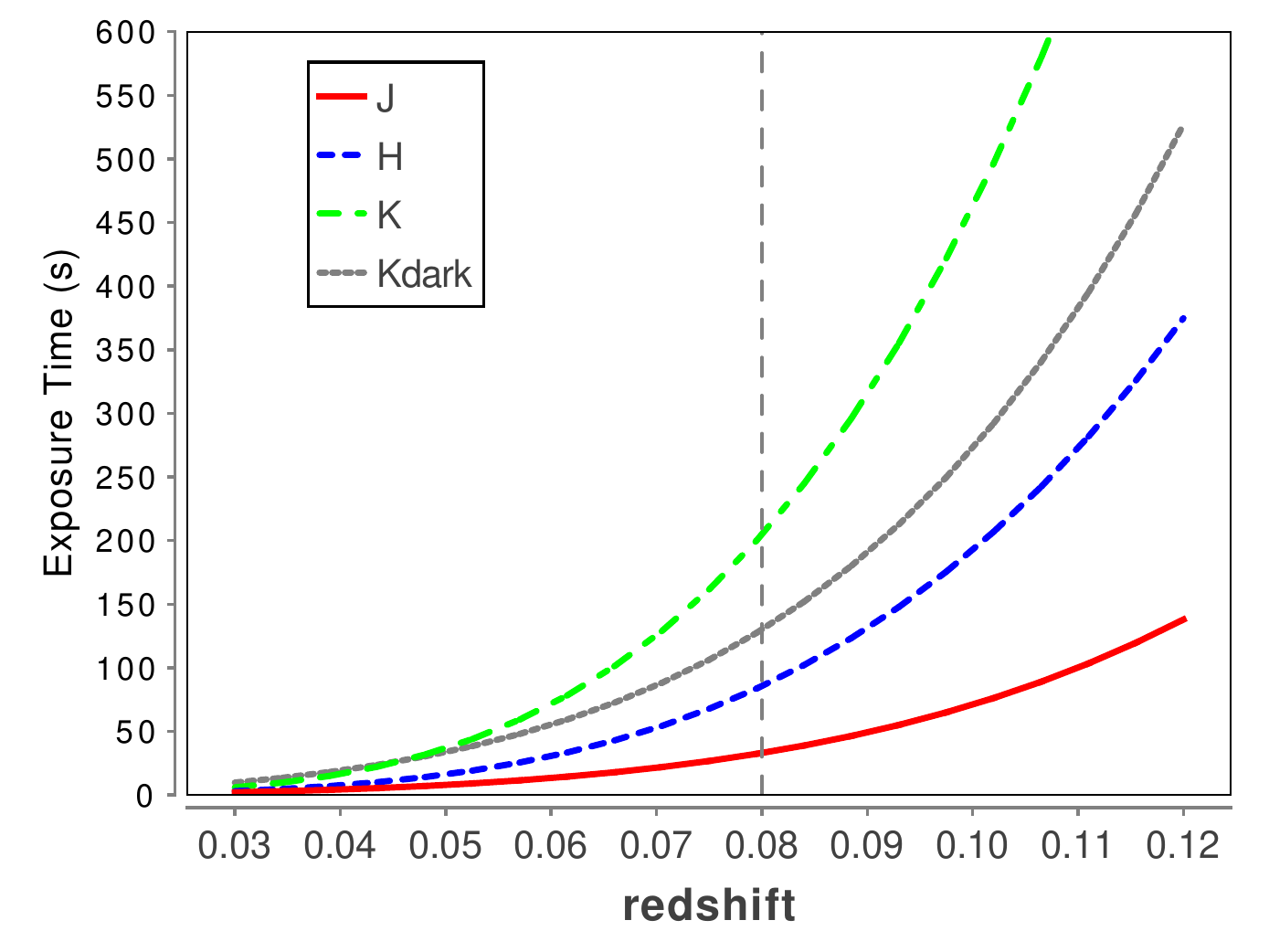}
\end{center}
\caption{The exposure time required to obtain $S/N=25$ at maximum light for a canonical
supernova using a 1-m telescope at Dome A with infrared bands $J$, $H$, $K$, and $K_{\rm dark}$.
The figure corresponds to a Galactic extinction of $E(B-V)=0.05$; exposure times 
are insensitive to Galactic dust at these red wavelengths.  \label{onem_nir:fig}}
\end{figure}

We assume that the 1-m telescope optics is diffraction-limited; at infrared wavelengths the instrument is an important contributor to the PSF.  The exposure times to give $S/N=25$ at peak brightness in infrared bands are shown in
Figure~\ref{onem_nir:fig}.  The exposure times needed to get both $J$ and $H$ bands together at $z=0.08$
is $\sim$120~s, similar to the time needed for spectroscopy.  $K$ band 
by itself would require $\sim$200~s whereas 
$K_{\rm dark}$,  with its low sky emission, requires $125$~s.

To summarize, the telescopes already planned for Dome A can execute a viable self-contained supernova survey. For example, data from a wide-field cadenced multi-band survey with AST3 can be processed locally to discover transients and use color and rise-time information to identify SNe Ia in the desired redshift range.  The 1-m telescope is triggered for targeted follow-up using
a camera capable of simultaneous observations of optical spectroscopy and infrared photometry
of the same field, via a dichroic.  The target is confirmed as SN Ia with the spectral
data and reobserved daily for the rest of the winter.  Deep references are taken
early in the next observing cycle when the 1-m is undersubscribed with a small list of live SN candidates. 
With co-located discovery and followup, non-interactive triggering and typing 
software that control telescope activities eliminate the need for live large data
transfer out of Dome A.  The full dataset can be physically retrieved for a posteriori
precision reprocessing.

Such instrumentation is also appropriate for the primary goal of 
site characterization, in its capability to measure the seeing distribution and wavelength-dependent sky emission out to $2.5$~$\mu$m.  For completeness, additional channels for the 1-m telescope can provide concurrent observation of an independent piece of sky with 
optical imaging and infrared spectroscopy.  The optical imaging provides important
PSF monitoring for the IFU data reduction.

\section{High-redshift Supernovae}
\label{highz:sec}
Only a few dozen high redshift ($z\gtrsim1$) supernovae are likely to exist on the Hubble diagram 
for the next several years.  With sufficiently accurate observations these play important roles 
by providing the potential for new discoveries and further constraining dark-energy 
parameters \cite{2003PhRvD..67h1303L}.  A large-aperture telescope at Dome A can serve three 
useful purposes: measuring lightcurves for a large sample of $z\sim1.5$ supernovae, obtaining 
good signal-to-noise photometry of high-redshift supernovae for discovery and triggering 
follow-up at other observatories, and exploring a small sample of $z\sim2.8$ supernovae through 
the Antarctic ``window'' of the $K_{\rm dark}$ band. 

The $\sim 5$ month night available for red to NIR observations at Dome A limits the coverage of the time-dilated light curves of high redshift supernovae.  A demanding light-curve requirement targets measurements of
the plateau phase so the period of two restframe 
months starting from explosion takes 5 observer months for a $z=1.5$ supernova.  
This means that only objects that explode at the beginning of the winter can have a 
full light curve, although the partial light curves would have much more detailed coverage than 
usual.  If the requirement were relaxed to obtain only the fall in magnitude from peak to 15 days 
after maximum  $\Delta m_{15}$, an important parameter correlated with absolute magnitude
\cite{1993ApJ...413L.105P}, this can be determined with 
one rest-frame month of data (pre-max data is essential to constrain the date of
maximum).  Such a survey could yield two and a half months worth of $z=1.5$ 
supernovae with a measurement of $\Delta m_{15}$ but no information tied to
homogeneity and heterogeneity at later epochs.
The size of the imager field-of-view could be chosen to ensure that sufficient numbers of supernovae explode in the narrow time window.  
The supernova rates at $z\sim 0.8$ give an expected 36 $\mbox{yr}^{-1} \mbox{deg}^{-2}$
per 0.1 redshift bin; 
rates at $z\sim1.5$ have 1-$\sigma$ uncertainties that range over a factor of five 
from 8--44 $\mbox{yr}^{-1} \mbox{deg}^{-2}$
per 0.1 redshift bin, and rates are observationally unconstrained beyond $z=1.7$
\cite{2008ApJ...673..981K,2008ApJ...681..462D}.  
As mentioned, lightcurves could be completed by feeding to mid-latitude telescopes. 

\begin{figure}
\begin{center}
\includegraphics[scale=0.48]{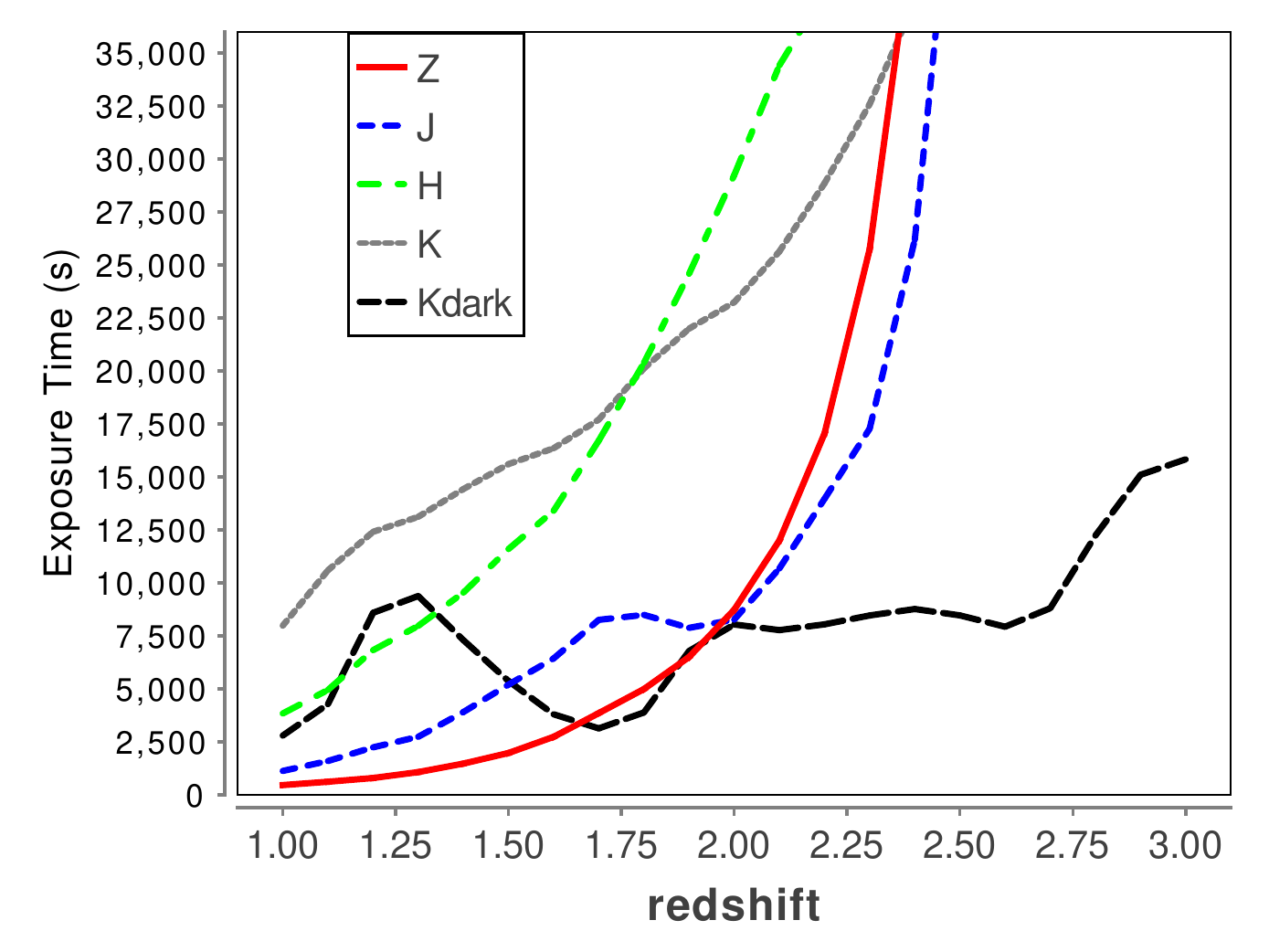}
\includegraphics[scale=0.48]{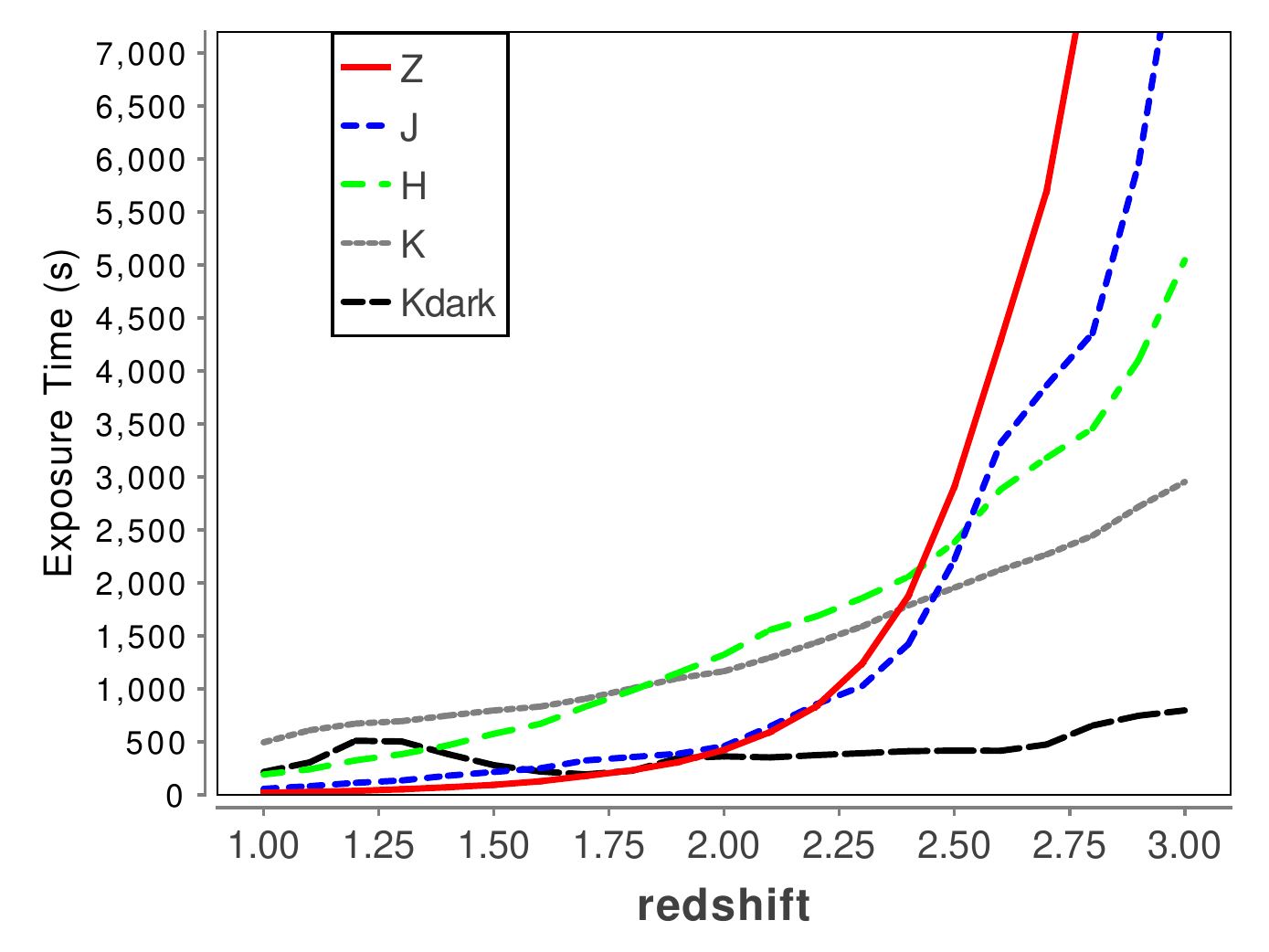}
\end{center}
\caption{The exposure times required to obtain $S/N=5$ five restframe days (left)
and ten restframe days (right) after explosion for a canonical 
supernova using an 8-m telescope at Dome A. Note the change in scale between the two plots. \label{deep:fig}}
\end{figure}

Obtaining large numbers of $z>0.8$ supernovae requires a small survey solid angle
compared to the low-redshift search, so 
we can select a search field with $\chi<1.5$ airmass and $E(B-V)\sim 0$.
Figure~\ref{deep:fig} shows for a range of redshifts the exposure times needed to get an early discovery of $S/N=5$ five and ten  
restframe days after explosion in different bands with an 8-m telescope.  With the high sky 
background, a small $10^{-4}$ flatfield uncertainty contributes non-negligibly to the overall error budget.  The $Z$ 
filter would be used with a thick fully-depleted CCD or HgCdTe whereas the infrared filters would be used with a HgCdTe detector.  Hours of integration time are necessary to detect objects 5-days after explosion out to $z \sim 2$ in standard bands while comparable exposures in $K_{\rm dark}$ can access supernovae out to $z=3$ and beyond (if there are indeed SNe Ia at these extreme redshifts).  Discovery 10 days after explosion requires less than an hour in all bands out to $z=2.5$ and tens of minutes in $K_{dark}$ out to $z=3$.  In this sky-background-noise dominated seeing-limited regime, the exposure times scale as $D^{-2}$
where $D$ is the primary mirror diameter.

\begin{figure}
\begin{center}
\includegraphics[scale=0.55]{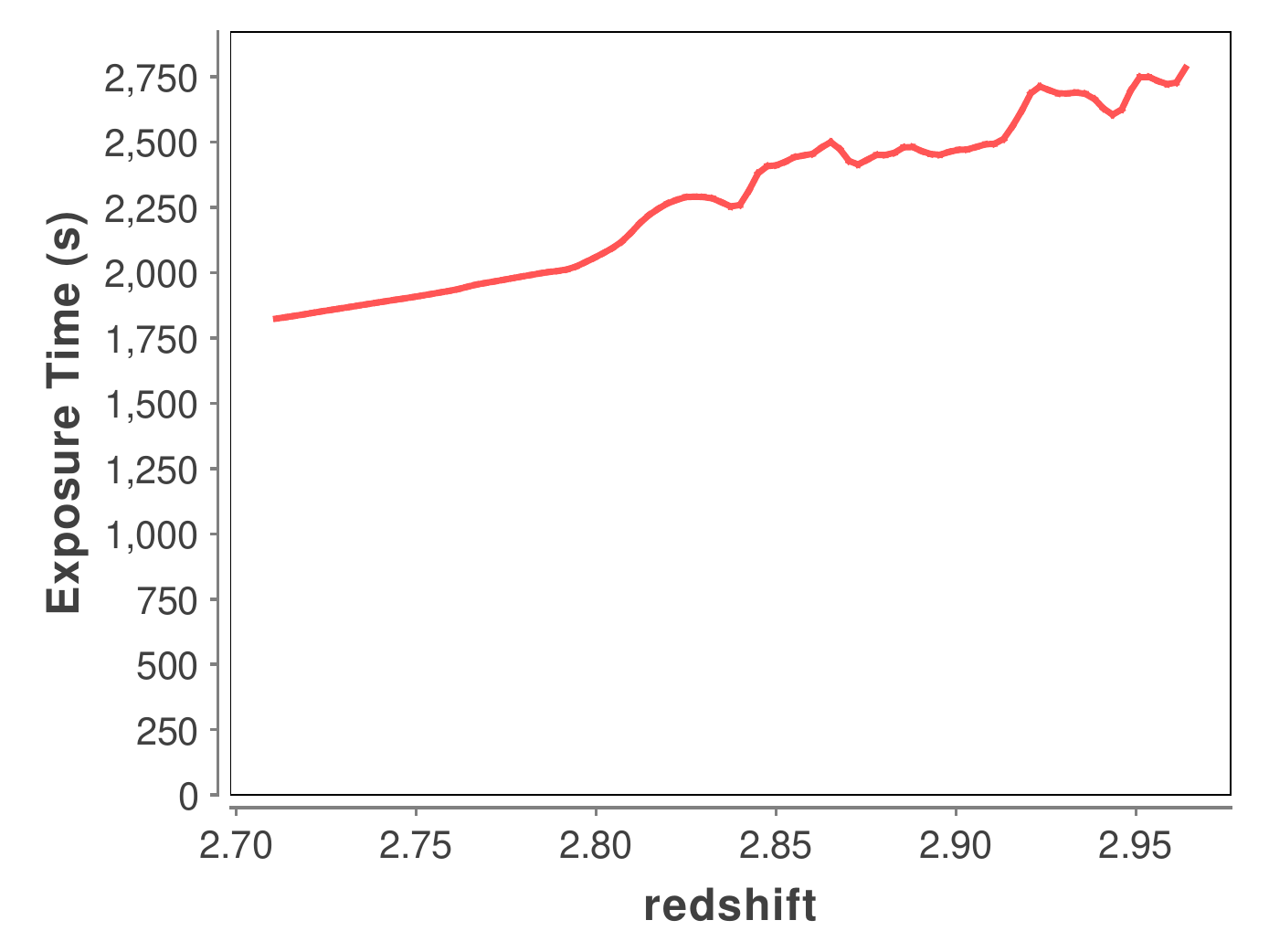}
\end{center}
\caption{The exposure time required to type a canonical 
supernova using an 8-m telescope at Dome A, with $S/N=5$ in a $\lambda/\delta \lambda=150$ element for the restframe 0.615 $\mu$m SiII feature. 
The restricted redshift range corresponds
to when the line is within the $K_{\rm dark}$ window. \label{deepType:fig}}
\end{figure}

The faint sky in $K_{\rm dark}$ also offers an interesting window in which we can observe
the SiII feature that defines the SN Ia class.  Figure~\ref{deepType:fig} shows
that half-hour exposures are needed to get $S/N=5$ in a $\lambda/\delta \lambda=150$ element for the restframe 0.615 $\mu$m feature at
peak brightness.  Dome A
thus provides an option for the typing of $2.75<z<2.95$ SNe Ia.  Typing at lower redshifts is restricted as the supernova spectrum is relatively featureless at longer wavelengths.

Depending on the requirements of the survey, high-redshift supernova observations from Dome A can 
provide a complete dataset for cosmology, or play an important role in a larger program.

\section{Conclusions}
\label{conclusions:sec}
We have considered how several telescopes planned for Dome A Antarctica can be used to
deliver SNe Ia science.  The first set of telescopes, AST3 and a 1-m pathfinder, can produce a photometric search and spectro-photometric survey for $z<0.08$ supernovae over an 8000 deg$^2$ region
of sky with low airmass and Galactic-dust absorption.  Depending on the goals of the survey, an instrument using slitless spectroscopy can provide a spectroscopic time series for all bright transients in the field.  
However, in this case the fine pixel sampling necessary to minimize sky background and the large survey solid angles lead to enormous numbers of detector pixels.

A next-generation large aperture telescope can take advantage of the excellent seeing and 
the $K_{\rm dark}$ observing window to discover  supernovae out to $z \sim 3$ and provide spectroscopic
typing in a specific redshift window.

On-site analysis and telescope control reduce communication needs out of Dome A. However, the more 
improvement in telemetry, the more that Antarctic advantages can be used in conjunction with transient surveys occurring elsewhere in the world.  Dome A discoveries can be promptly announced for observations at other observatories, and external discoveries can be followed at Dome A.

More information on site performance would enable better treatment beyond several simplifying assumptions in our calculations.  We did not include host-galaxy background nor host-galaxy extinction.  Our calculations are based on the fixed median seeing and not the full distribution (which can drop to 0.1").
We don't simulate the full range of SNe Ia but base our calculations on an average supernova.  Exposure time estimates are based on $z=0.08$, the high end of the targeted redshift range.  We do not offer precise numbers on how far into twilight supernova observations are possible, i.e.\ the number of hours per night (we assume 16 hours) and the fraction of the year (we assume five months) that can be spent on the survey.  Patterns of suspended observations due to weather are not considered.  As more information on the site is collected, we can add more realism into our projections. 
Nevertheless, these first calculations show that interesting SN~Ia science
can be done at Dome A.

A number of other science drivers are possible as well.  
Mapping the peculiar velocity field at $z<0.03$ using SNe Ia is an interesting
probe of cosmology \cite{1997ApJ...488L...1R,2007arXiv0705.0368W}.  With a SNe Ia rate of 
$\sim 0.006$ $\mbox{deg}^{-2}\mbox{yr}^{-1}$, around ten years are required to build significant
statistics in the restricted field available at Dome A.  

Core-collapse supernovae, in particular Type IIP's, are of interest 
for cosmology \cite{2006ApJ...645..841N,2009ApJ...694.1067P,2009ApJ...696.1176J}.  Exposure times can be tuned to discover these
fainter objects and the continuous observing allows monitoring for the UV shock breakout \cite{2007ApJ...666.1093Q}.  The late-time
spectroscopy needed to determine velocities of the ejecta to standardize them as distance indicators does limit the time window that these supernovae can be discovered and followed
at Dome A.

We have not considered the possibility of using adaptive optics at Dome A, although the greater 
isoplanatic angle and coherence time make this of interest.  Such a capability
can aid in supernova typing, but it is uncertain how it can achieve precision photometry of point sources on top of an underlying host galaxy over a large field of view.

Finally, the interest in an instrument that provides simultaneous optical IFU spectroscopy and NIR imaging of the same field is not unique to the site.  Such a camera lessens the need for multi-telescope coordination in building a large set of pan-chromatic supernova data.

There remain considerable areas of interest to explore in Antarctic astronomy, 
supernova surveys, and their intersection.  This article has presented a first look 
at some of the prospects and issues.  With the ongoing development at Dome A and the 
near-term installation of wide-field and 1-m class telescopes, these topics are worth 
pursuing further.

\section{Acknowledgements}
This work was supported by the Director, Office of Science, Office of High Energy Physics, 
of the U.S.\ Department of Energy under Contract No. DE-AC02-05CH11231 and
DE-AC02-07CH11359.  LW is supported partially by an NSF grant AST 0708873.

\bibliographystyle{elsarticle-num}

\begin{thebibliography}{10}
\expandafter\ifx\csname url\endcsname\relax
  \def\url#1{\texttt{#1}}\fi
\expandafter\ifx\csname urlprefix\endcsname\relax\def\urlprefix{URL }\fi
\expandafter\ifx\csname href\endcsname\relax
  \def\href#1#2{#2} \def\path#1{#1}\fi

\bibitem{2009astro2010S.308W}
L.~{Wang} et al., {Astronomy on Antarctic Plateau}, Astro2010: The Astronomy and Astrophysics Decadal
  Survey, Vol. 2010 of Astronomy, 2009, pp. 308.

\bibitem{2009PASP..121..976S}
W.~{Saunders} et al,  \pasp, 121, 2009, 976--992.

\bibitem{1999ApJ...527.1009P} Phillips, A., Burton, 
M.~G., Ashley, M.~C.~B., Storey, J.~W.~V., Lloyd, J.~P., Harper, D.~A., 
\& Bally, J.\ 1999, \apj, 527, 1009 

\bibitem{2005PASA...22..199B}
M.~G. {Burton} et al., Publications of the
  Astronomical Society of Australia 22, 2005, 199--235.
  
 \bibitem{2009PASP..121..174Y}
H.~{Yang} et al., \pasp, 121, 2009, 174--184.

\bibitem{2006PhRvD..74j3518L}
E.~V. {Linder}, 
  \prd, 74~(10), 2006, 103518.

\bibitem{2002SPIE.4836...61A}
G.~{Aldering} et al., in: {J.~A.~Tyson \& S.~Wolff} (Ed.), Society of
  Photo-Optical Instrumentation Engineers (SPIE) Conference Series, Vol. 4836
  of Society of Photo-Optical Instrumentation Engineers (SPIE) Conference
  Series, 2002, pp. 61--72.

\bibitem{2008ApJ...682..262D}
B.~{Dilday} et al., \apj, 682, 2008, 262--282.

\bibitem{1998ApJ...500..525S}
D.~J. {Schlegel}, D.~P. {Finkbeiner}, M.~{Davis}, \apj, 500, 1998,  525.

\bibitem{2007ApJ...663.1187H}
E.~Y. {Hsiao} et al., \apj, 663, 2007,  1187--1200.

\bibitem{2002AJ....123..745R}
D.~{Richardson}, D.~{Branch}, D.~{Casebeer}, J.~{Millard}, R.~C. {Thomas},
  E.~{Baron}, \aj, 123, 2002, 745--752.

\bibitem{2002SPIE.4835..146A}
G.~{Aldering} et al., in: {A.~M.~Dressler} (Ed.), Society of Photo-Optical
  Instrumentation Engineers (SPIE) Conference Series, Vol. 4835 of Society of
  Photo-Optical Instrumentation Engineers (SPIE) Conference Series, 2002, pp.
  146--157.

\bibitem{2009A&A...500L..17B}
S.~{Bailey} et al., \aap, 500, 2009, L17--L20.

\bibitem{2007ApJ...671.1084S}
M.~{Strovink}, \apj, 671, 2007, 1084--1097.

\bibitem{2003LNP...635..203H}
P.~{H{\"o}flich}, C.~{Gerardy}, E.~{Linder}, {et al.}, in: {D.~Alloin \& W.~Gieren} (Ed.), Stellar
  Candles for the Extragalactic Distance Scale, Vol. 635 of Lecture Notes in
  Physics, Berlin Springer Verlag, 2003, pp. 203--227.

\bibitem{2004ApJ...602L..81K}
K.~{Krisciunas}, M.~M. {Phillips}, N.~B. {Suntzeff}, \apjl, 602, 2004, L81--L84.

\bibitem{2008ApJ...689..377W}
W.~M. {Wood-Vasey} et al., \apj, 689, 2008, 377--390.

\bibitem{2006ApJ...649..939K}
D.~{Kasen},  \apj, 649, 2006, 939--953.

\bibitem{2003PhRvD..67h1303L}
E.~V. {Linder}, D.~{Huterer}, \prd, 67~(8), 2003, 081303.

\bibitem{1993ApJ...413L.105P}
M.~M. {Phillips}, \apjl, 413, 1993, L105--L108.

\bibitem{2008ApJ...673..981K}
N.~{Kuznetsova} et al., \apj, 673, 2008, 981--998.

\bibitem{2008ApJ...681..462D}
T.~{Dahlen}, L.~{Strolger}, A.~G. {Riess},  \apj, 681, 2008, 462--469.

\bibitem{1997ApJ...488L...1R}
A.~G. {Riess}, M.~{Davis}, J.~{Baker}, R.~P. {Kirshner}, \apjl,
  488, 1997, L1.

\bibitem{2007arXiv0705.0368W}
L.~{Wang}, ArXiv e-prints arXiv:0705.0368.

\bibitem{2006ApJ...645..841N}
P.~{Nugent} et al., \apj, 645, 2006, 841--850.

\bibitem{2009ApJ...694.1067P}
D.~{Poznanski} et al., \apj 694, 2009, 1067--1079.

\bibitem{2009ApJ...696.1176J}
M.~I. {Jones} et al., \apj, 696,
2009, 1176--1194.

\bibitem{2007ApJ...666.1093Q}
R.~M. {Quimby}, J.~C. {Wheeler}, P.~{H{\"o}flich}, C.~W. {Akerlof}, P.~J.
  {Brown}, E.~S. {Rykoff}, \apj, 666, 2007, 1093--1107.

\end{thebibliography}

\end{document}